\documentclass[a4paper,11pt]{article}
\pdfoutput=1 % if your are submitting a pdflatex (i.e. if you have
\usepackage{jcappub} % for details on the use of the package, please
\usepackage[T1]{fontenc} % if needed

\usepackage{slashed}
\usepackage{bm}
\usepackage{latexsym,amssymb,amsmath,float,url,mathrsfs}
\usepackage{latexsym}
\usepackage{graphicx}
\usepackage{epstopdf}
\usepackage{amsmath}
\usepackage{comment}
\usepackage{subfigure}
\usepackage{natbib}
\usepackage{hyperref}
\usepackage{pifont}
\usepackage{blindtext}
\usepackage{adjustbox}
\usepackage{multirow}
\usepackage{tabularx}
\usepackage[table]{xcolor}
\usepackage{orcidlink}
\usepackage{tikz}
\usetikzlibrary{tikzmark}
\usepackage{fontawesome}

\usepackage{blkarray}
\usepackage{graphicx}
\usepackage{amsfonts}
\usepackage{soul}
\usepackage{amssymb}
\usepackage{amsmath}
\usepackage{cancel}
\usepackage{tcolorbox}

\usepackage{graphics,appendix,afterpage,makecell} 

\definecolor{oucrimsonred}{rgb}{0.6, 0.0, 0.0}
\definecolor{persianblue}{rgb}{0.11, 0.22, 0.73}
\definecolor{forestgreen}{rgb}{0.13,0.35,0.13}
\definecolor{lightgray}{rgb}{0.83, 0.83, 0.83}
 \hypersetup{colorlinks, citecolor=oucrimsonred, linkcolor=black, urlcolor=oucrimsonred}
\definecolor{cornellred}{rgb}{0.7, 0.11, 0.11}
\definecolor{navyblue}{rgb}{0.0, 0.0, 0.5}
\definecolor{amethyst}{rgb}{0.6, 0.4, 0.8}
\definecolor{yellow}{rgb}{1.0, 1.0, 0.0}
\definecolor{firebrick}{rgb}{0.7, 0.13, 0.13}
\definecolor{tangerineyellow}{rgb}{1.0, 0.8, 0.0}
\definecolor{deepfuchsia}{rgb}{0.76, 0.33, 0.76}
\definecolor{amber}{rgb}{1.0, 0.75, 0.0}
\definecolor{VioletRed4}{rgb}{0.55, 0.13, .32}
\definecolor{indiagreen}{rgb}{0.07, 0.53, 0.03}
\definecolor{VioletRed4}{rgb}{0.55, 0.13, .32}
\newcommand{\be}{\begin{equation}}
\newcommand{\ee}{\end{equation}}
\newcommand{\bea}{\begin{equation} \begin{aligned}}
\newcommand{\eea}{\end{aligned} \end{equation}}

\definecolor{oucrimsonred}{rgb}{0.6, 0.0, 0.0}
\newcommand\vertarrowbox[3][6ex]{%
  \begin{array}[t]{@{}c@{}} #2 \\
  \left\uparrow\vcenter{\hrule height #1}\right.\kern-\nulldelimiterspace\\
  \makebox[0pt]{\scriptsize#3}
  \end{array}%
}

\definecolor{violachiaro}{rgb}{1,0.6,1}

\definecolor{gbcolor}{rgb}{.43,.22,.12}
 
\definecolor{gbcolor2}{rgb}{.9,.2,.6}
\definecolor{gbcolor3}{rgb}{.3,.2,.6}

\definecolor{verdechiaro}{rgb}{0.6,1,0.6}
\definecolor{giallochiaro}{rgb}{1,1,0.6}
\definecolor{bluscuro}{rgb}{0.15, 0.2, 0.9}
\definecolor{verdes}{rgb}{0.1, 0.5, 0.1}%
\definecolor{tangerineyellow}{rgb}{1.0, 0.8, 0.0}

\definecolor{americanrose}{rgb}{1.0, 0.01, 0.24}
\definecolor{cobalt}{rgb}{0.0, 0.28, 0.67}
\definecolor{brandeisblue}{rgb}{0.0, 0.44, 1.0}
\definecolor{mycolor}{rgb}{0.0, 0.0, 0.5}%navyblue
\definecolor{oxfordblue}{rgb}{0.0, 0.13, 0.28}
\definecolor{azure}{rgb}{0.0, 0.5, 1.0}
\definecolor{turquoiseblue}{rgb}{0.0, 1.0, 0.94}
\newtcolorbox{mynewbox}[1]{colback=white!5!white,colframe=azure!75!black,fonttitle=\bfseries,title=#1}
\newtcolorbox{mybox}{colback=mycolor!5!white,colframe=azure!75!black}
\newtcolorbox{mynamedbox}[1]{colback=mycolor!5!white,colframe=azure!75!black,title=#1}
\definecolor{venetianred}{rgb}{0.78, 0.03, 0.08}
\newtcolorbox{mynamedbox1}[1]{colback=venetianred!5!white,colframe=venetianred!80!black,title=#1}
\newtcolorbox{mynamedbox2}[1]{colback=azure!5!white,colframe=azure!80!black,title=#1}

\definecolor{verdes}{rgb}{0.1, 0.5, 0.1}%
\definecolor{cornellred}{rgb}{0.7, 0.11, 0.11}

\definecolor{VioletRed4}{rgb}{0.55, 0.13, .32}

\hypersetup{
     colorlinks   = true,
     citecolor    = violet,
     urlcolor     = violet,
     linkcolor    = violet}

\definecolor{ForestGreen}{rgb}{0.13, 0.55, 0.13}

\definecolor{rossocorsa}{rgb}{0.83, 0.0, 0.0}

\newcommand{\uniroma}{Dipartimento di Fisica, ``Sapienza'' Universit\`a di Roma, Piazzale Aldo Moro 5, 00185, Roma, Italy}
\newcommand{\infn}{INFN sezione di Roma, Piazzale Aldo Moro 5, 00185, Roma, Italy}

\title{Novel tests of gravity using nano-Hertz\\ stochastic gravitational-wave background signals}

\author[a,b]{Enrico Cannizzaro}
\author[a,b]{Gabriele Franciolini}
\author[a,b]{Paolo Pani}

\affiliation[a]{\uniroma}
\affiliation[b]{\infn}

\emailAdd{enrico.cannizzaro@uniroma1.it}
\emailAdd{gabriele.franciolini@uniroma1.it}
\emailAdd{paolo.pani@uniroma1.it}

\abstract{Gravity theories that modify General Relativity in the slow-motion regime can introduce nonperturbative corrections to the stochastic gravitational-wave background~(SGWB) from supermassive black-hole binaries in the nano-Hertz band, 
while not affecting the quadrupolar nature of the gravitational-wave radiation and remaining perturbative in the highly-relativistic regime, as to satisfy current post-Newtonian~(PN) constraints. We present a model-agnostic formalism to map such theories into a modified tilt for the SGWB spectrum, showing that negative PN corrections (in particular -2PN) can alleviate the tension in the recent pulsar-timing-array data if the detected SGWB is interpreted as arising from supermassive binaries. Despite being preliminary, current data have already strong constraining power, for example they set a novel (conservative) upper bound on theories with time-varying Newton's constant {(a $-4$PN correction)} at least at the level of $\Dot{G}/G \lesssim 10^{-5} \text{yr}^{-1}$ for redshift $z=[0.1\div1]$. We also show that NANOGrav data are best fitted by a broken power-law interpolating between a dominant -2PN or -3PN modification at low frequency, and the standard general-relativity scaling at high frequency. Nonetheless, a modified gravity explanation should be confronted with binary eccentricity, environmental effects, nonastrophysical origins of the signal, and scrutinized against statistical uncertainties. These novel tests of gravity will soon become more stringent when combining all pulsar-timing-array facilities and when collecting more~data.}

\begin{document}
\maketitle
\flushbottom

\section{Introduction} 
Pulsar timing arrays~(PTAs) offer a unique way to probe gravitational-wave~(GW) astrophysics at the nano-Hertz~(nHz) scale. In 2020, the NANOGrav collaboration first reported evidence in their 12.5 year dataset~\cite{NANOGrav:2020bcs} for a common spectrum of a stochastic nature, which provided the first hint of a stochastic gravitational wave background~(SGWB) signature.  However, in these data, there was no statistical evidence for Hellings-Down~(HD) correlation pattern, necessary to interpret the signal as a quadrupolar GW background. 
Remarkably, the more recent PTA data released in 2023 by the NANOGrav~\cite{NG15-SGWB,NG15-pulsars}, EPTA (in combination with InPTA)\,\cite{EPTA2-SGWB,EPTA2-pulsars,EPTA2-SMBHB-NP}, PPTA\,\cite{PPTA3-SGWB,PPTA3-pulsars,PPTA3-SMBHB} and CPTA\,\cite{CPTA-SGWB} collaborations, found evidence for a HD angular correlation,  typical of an homogeneous spin-2 GW background and consistent with the quadrupolar nature of GWs in General Relativity (GR)~\cite{1983ApJ...265L..39H}.

The latest NANOGrav $15\,{\rm yr}$ (henceforth NANOGrav15) data found evidence for a smooth power law, $\Omega_{\rm GW}\propto f^{(1.6, 2.3)}$ at $1\sigma$. It is well known that a SGWB sourced by supermassive black hole (SMBH) binaries would be characterized by a scaling law $\Omega_{\rm GW} \propto f^{2/3}$~\cite{Phinney:2001di}, which is currently disfavoured at $2\sigma$ by the NANOGrav15 data\,\cite{NG15-SMBHB,NG15-NP}. Nevertheless, environmental and statistical effects may play a relevant role, and lead to a steeper scaling ~\cite{Sesana:2008mz,2010arXiv1002.0584K,Kelley:2016gse,Perrodin:2017bxr,Ellis:2023owy,NG15-SMBHB,NG15-NP,Ghoshal:2023fhh, Bonetti:2017lnj}.
Even though with current data it is not possible to rule out a cosmological origin for the observed signal -- such as coming from first-order phase transitions~\cite{NANOGrav:2021flc,Xue:2021gyq,Nakai:2020oit,DiBari:2021dri,Sakharov:2021dim,Li:2021qer,Ashoorioon:2022raz,Benetti:2021uea,Barir:2022kzo,Hindmarsh:2022awe,Gouttenoire:2023naa,Baldes:2023fsp,Li:2023bxy}, cosmic strings and domain walls~\cite{Ellis:2020ena,Datta:2020bht,Samanta:2020cdk,Buchmuller:2020lbh,Blasi:2020mfx,Ramazanov:2021eya,Babichev:2021uvl,Gorghetto:2021fsn,Buchmuller:2021mbb,Blanco-Pillado:2021ygr,Ferreira:2022zzo,An:2023idh,Qiu:2023wbs,Zeng:2023jut,King:2023cgv,Babichev:2023pbf, Kitajima:2023cek,Barman:2023fad}, or scalar-induced GWs generated from primordial fluctuations~\cite{Vaskonen:2020lbd,DeLuca:2020agl,Bhaumik:2020dor,Inomata:2020xad,Kohri:2020qqd,Domenech:2020ers,Vagnozzi:2020gtf,Namba:2020kij,Sugiyama:2020roc,Zhou:2020kkf,Lin:2021vwc,Rezazadeh:2021clf,Kawasaki:2021ycf,Ahmed:2021ucx,Yi:2022ymw,Yi:2022anu,Dandoy:2023jot,Zhao:2023xnh,Ferrante:2023bgz,Cai:2023uhc,Franciolini:2023pbf,Balaji:2023ehk, Liu:2023ymk} (see also~\cite{Franciolini:2023wjm,madge2023primordial,Figueroa:2023zhu, Garcia-Bellido:2023ser, Murai:2023gkv, Konoplya:2023fmh, EPTA:2023xiy}) --
the SMBH hypothesis remains the leading explanation. Assuming the signal had astrophysical origin, one can greatly constrain the scenario and the physics governing SMBH binaries.

In principle, also modifications to GR (see, e.g.,~\cite{Will:2014kxa,Yunes:2013dva,Berti:2015itd} for some reviews) may play a crucial role and lead to a different prediction than the standard scaling expected for SMBH binaries {(including correlation patterns deviating from HD \cite{Chamberlin:2011ev,Qin:2020hfy,Liang:2023ary,Bernardo:2023pwt})}. Indeed, many known theories beyond GR induce new effects at negative post-Newtonian~(PN) orders. Examples include dipole radiation in scalar-tensor theories at $-1\text{PN}$, or the effects of extra-dimensions or a time-varying Newton constant,  both at $-4\text{PN}$ (see e.g.~\cite{Berti:2015itd,Yunes:2016jcc,Barausse:2016eii}). Given that BH binaries in the PTA band have extremely large orbital separation during the early-inspiral phase, one could expect negative PN modifications to GR in this regime to play a much more relevant role than in the coalescence phase typically explored by ground- and space-based detectors~\cite{LIGOScientific:2021sio,Maggiore:2019uih,Branchesi:2023mws,Barausse:2020rsu,LISA:2022kgy}.

In this work, 
we focus on modified GR theories preserving the quadrupolar nature of the gravitational signal, i.e that would be compatible with the HD correlation pattern and with LIGO-Virgo-KAGRA constraints on the GW polarizations~\cite{LIGOScientific:2021sio}. 
For example, widely considered scalar-tensor theories introduce new dissipative degrees of freedom (e.g., scalar fields) modifying the dynamics of the binary (through extra energy fluxes) but without coupling to matter nor to the pulsar signal~\cite{Berti:2015itd,Yunes:2016jcc}. These theories propagate GWs with only the two standard GR polarizations~\cite{Iyonaga:2018vnu,Wagle:2019mdq} but introduce negative PN corrections in the GW signal. 
We therefore analyze the impact of generic effects at negative PN orders in light of the recent results of the PTA collaborations. We show that the recently detected SGWB allows for novel tests of GR. On the one hand, slow-motion modifications to GR can alleviate the current tension in PTA data and, on the other hand, we can use current data to place stringent upper bounds on putative negative-PN modifications, e.g. theories predicting a time-varying Newton's constant, $G(t)$, in an unconstrained region of their parameter space~\cite{units}.

\section{SGWB spectrum in modified gravity theories} 
The Hamiltonian of a binary, corresponding to the centre of mass binding energy reads, at the Newtonian level,
\begin{equation}
    E=\mu v^2/2-G m_1 m_2/r\,, \label{eq:Hamiltonian}
\end{equation}
where $m_1, m_2$ are the masses of the objects, $r$ is the relative orbital separation, $v$ is the relative orbital velocity, and $\mu$ is the reduced mass.
Here we adopt a theory agnostic approach and consider model-independent modifications to GR. In particular, we consider \textit{dissipative} corrections to the rate of change of the binding energy,  at generic PN order\footnote{One can also include a similar parametrization for the \textit{conservative} corrections to the binding energy of the binary~\cite{Yunes:2009ke,Chatziioannou:2012rf}. For simplicity, here we focus only on dissipative corrections.}:
\begin{equation}
\label{eq:Beq}
    \Dot{E}=\Dot{E}_{\rm GR}\Big[1+ B \Big (\frac{G m}{r}\Big)^q\Big]\,,
\end{equation}
where $m=m_1+m_2$ is the total mass and $\Dot{E}_{\rm GR}=\frac{32}{5}G\mu^2\omega^6 r^4$ is the standard GR energy flux. This approach was widely considered in the literature in the contest of the parametrized post-Einsteinian~(ppE) framework, wherein the parameter $q$ is theory-dependent and accounts for the dominant correction, while the parameter $B$ can be mapped to the coupling constants of a given theory~\cite{Yunes:2009ke, Chatziioannou:2012rf, Barausse:2016eii,Maselli:2016ekw}.
Nevertheless, in the ppE formalism these corrections are always assumed to be much smaller than the GR term, and hence treated perturbatively. Here we remove this assumption, as we are interested in a seldom explored regime where these terms might dominate over the GR ones at the nHz scale.  To compute the impact of this modification in the energy spectrum of GWs, we can simply express the latter as
\begin{equation}
\label{eq:dEdw}
    \frac{d E_{\rm GW}}{d\omega}=\frac{\Dot{E}_{\rm GR}}{\Dot{\omega}}\, .
\end{equation}
As we focused on purely dissipative corrections to GR, the binding energy is not modified, and one can therefore assume  the standard Kepler's law relating the orbital radius to the frequency at the leading order, $r=(G m /\omega^2)^{1/3}$.
Using the latter, and combining Eqs.~\eqref{eq:Hamiltonian} and~\eqref{eq:Beq}, we can compute the rate of change of the orbital frequency,
\begin{equation}
    \Dot{\omega}=\frac{96}{5}\mu G^{5/3}m^{2/3}\omega^{11/3}\Big[1+ B \Big (\frac{G m}{r}\Big)^q\Big]\,.
\end{equation}
Finally, the modified GW spectrum reads
\begin{equation}
\label{eq:dEdfB}
    \frac{d E_{\rm GW}}{d\omega}=-\frac{\frac{1}{3}G^{2/3}\mu m^{2/3} \omega^{-1/3}}{1+ B (G m \omega)^{2q/3}}\, .
\end{equation}
In the $B=0$ limit this equation coincides with the standard GR one, where the only dissipation channel is ordinary GW emission, and leads to the standard scaling $\Omega_{\rm GW} \propto f dE_{\rm GW}/df\sim f^{2/3}$, where $f= \omega/\pi$ is the GW frequency. On the other hand, if 
\begin{equation}
   B (G m \pi f)^{2q/3}\gg 1\,,
   \label{eq:inequality}
\end{equation}
the modified scaling reads
\begin{equation}
    \Omega_{\rm GW} \propto f^{\frac{2}{3}(1-q)}\, .
\end{equation}
Modifications of the energy flux for \textit{negative} PN orders ($q<0$) lead to a steeper scaling, as expected. In particular, a dominant $-2 \text{PN}$ correction ($q=-2$) would give $\Omega_{\rm GW} \propto f^2$, compatible with the peak of NANOGrav15 posteriors for the tilt~\cite{NG15-SGWB}, as we shall also discuss later on.
Taking $q=-2$ as an example, from Eq.~\eqref{eq:inequality} one has 
\begin{equation}
    B\gg 4\times 10^{-7} \left(\frac{m}{10^9 M_\odot}\right)^{4/3}\left(\frac{f}{\rm nHz}\right)^{4/3}\,. \label{eq:inequality2}
\end{equation}
On the other hand, the requirement that the same correction remains perturbative during the inspiral of a coalescing binary imposes $B\ll(Gm \omega_{\rm ISCO})^{-2q/3}\sim 0.03$ (for $q=-2$ and where $\omega_{\rm ISCO}$ is the reference frequency of the innermost stable circular orbit), regardless of the binary mass. Thus, there is a wide range in which the coupling $B$ satisfies the inequality~\eqref{eq:inequality2} while introducing a small correction in the coalescence phase.
This range becomes even wider for more negative PN corrections.

Note that this general formalism includes several classes of gravity theories~\cite{Yunes:2016jcc}, but also several environmental effects that provide dissipation mechanisms at negative PN orders~\cite{Barausse:2014tra,Barausse:2014pra,Cardoso:2019rou}, 
leading to a steeper scaling~\cite{2010arXiv1002.0584K,Sampson:2015ada,Ghoshal:2023fhh}. Examples include accretion and dynamical friction ($q=-5.5$), stellar scattering ($q=-5$), and interaction with circumbinary gas ($q=-3.5$). Finally, large eccentricity ($e\gtrsim0.6$) also provides a steeper scaling~\cite{Enoki:2006kj,Chen:2016zyo} which can be fitted by multiple power-law phases with $q\approx-7.1$ and $q\approx-2.8$.

\section{The case of varying G}
Before delving into an actual confrontation with PTA data in the next section, let us discuss the strong constraining power of current data through a heuristic argument and considering a specific example.
Alternatives to GR can violate the strong equivalence principle and in some case break local invariance~\cite{Will:2014kxa}, while remaining (mostly) quadrupolar in nature. Some of them predict a spacetime variation of the effective Newton constant, for example mediated by a scalar field on cosmological scales~\cite{Fujii:2003pa} or in the presence of energy leakage into small extra dimensions~\cite{Deffayet:2007kf,Yagi:2011yu}. Here we apply our formalism to the simplest and most studied case, promoting $G$ to a function of time, $G(t)$, with $G_0$ being its present value~\cite{Damour:1988zz,Yunes:2009bv}.
Several constraints to the first derivative of the Newton constant, $\Dot{G}$, have been considered at different scales. At cosmological scales, bounds come from Big Bang nucleosynthesis~\cite{Copi:2003xd,Bambi:2005fi,Alvey:2019ctk} and cosmic microwave background~\cite{Wu:2009zb}, which estimated $\Dot{G}/G_0\lesssim 10^{-12}\,\text{yr}^{-1}$, while in the solar system the most stringent constraint comes from a detailed analysis of Mercury's orbits, yielding $\Dot{G}/G_0\lesssim 10^{-14}\,\text{yr}^{-1}$~\cite{2018NatCo...9..289G}. Nevertheless, the former constraint assumes a linear scaling of $G(t)$ throughout the entire cosmic history, while the latter is obtained in the solar system at zero redshift. On the other hand, GWs offer a unique probe of \textit{local} variation $\Dot{G}$ at intermediate epochs~\cite{Yunes:2009bv,Barbieri:2022zge},  being thus complementary to the aforementioned constraints.   Depending on the specific GW source, one can place constraints at different redshift $z$. In particular, PTAs are sensible to systems at $z \approx [0.1 \div 1]$ \cite{NG15-SMBHB}, and hence they can provide complementary constraints to the low-redshift ones which were placed using binary pulsars~\cite{Damour:1988zz,1994ApJ...428..713K,Thorsett:1996fr} and LIGO~\cite{LIGOScientific:2021sio}. On the other hand, the future space mission LISA will be able to provide bounds at similar redshifts~\cite{Yunes:2009bv,Chamberlain:2017fjl,Perkins:2020tra,Barbieri:2022zge}.
Here we show that using the recent PTA detection of the SGWB it is possible to place competitive bounds already with current data. The GW emission power in these theories depends on the time-varying Newton constant,
\begin{equation}
    \Dot{E}_{\rm GW}=\frac{32}{5 G(t)}\Big[{\pi \mathcal{M}_c}G(t)f(t) \Big]^{10/3}\, ,
\end{equation}
where $\mathcal{M}_c$ is the binary chirp mass. Using Eq.~\eqref{eq:Hamiltonian} (with $G\to G(t)$) and the balance law, one can compute an expression for the secular variation of the GW frequency valid at all orders in $\Dot{G}$~\cite{Barbieri:2022zge}: 
\begin{equation}
\label{eq:dotfvaryingG}
    \Dot{f}=\frac{96 \pi^{8/3}}{5}G^{5/3}\mathcal{M}_c^{5/3}f^{11/3}-\Big(\frac{\Dot{G}}{G} \Big)f\, .
\end{equation}
The first term above is the standard GR one, while the second term is the modification due to the variation of $G$.
Using the above relations, one can straightforwardly compute Eq.~\eqref{eq:dEdw}. If the first term dominates over the second one in Eq.~\eqref{eq:dotfvaryingG}, one obtains the standard scaling $\Omega_{\rm GW} \propto f^{2/3}$, while in the opposite regime, i.e.\ if $\Dot{G}$ is large enough, one obtains the scaling
\begin{equation}
    \Omega_{\rm GW} \propto f^{10/3}\, , \label{eq:spectrumGdot}
\end{equation}
which corresponds to $q=-4$, i.e.\ to a $-4{\rm PN}$ order effect, in agreement with the ppE mapping of this correction in the perturbative regime~\cite{Yunes:2009bv,Yunes:2016jcc}. Hence, theories with a time-varying $G$ can be directly mapped into our generic framework. The obtained scaling~\eqref{eq:spectrumGdot} deviates more than $3 \sigma$ from the recent NANOGrav15 measurement, which can in turn be used to  place constraints on $\Dot{G}$. Assuming for simplicity that all binaries have same redshift and equal masses, $m_1=m_2=m/2$, one finds that the spectrum~\eqref{eq:spectrumGdot} is obtained if 
\begin{equation}
    \Dot{G}/G_0 \gg 4.6 \times 10^{-9} \text{yr}^{-1} \left(\frac{f}{\rm nHz}\right)^{8/3}\left(\frac{m}{10^9 M_\odot}\right)^{5/3}\, , \label{eq:bound}
\end{equation}
which is therefore excluded by PTA measurements.

\begin{figure}[t]
\centering
\includegraphics[width=0.6\textwidth]{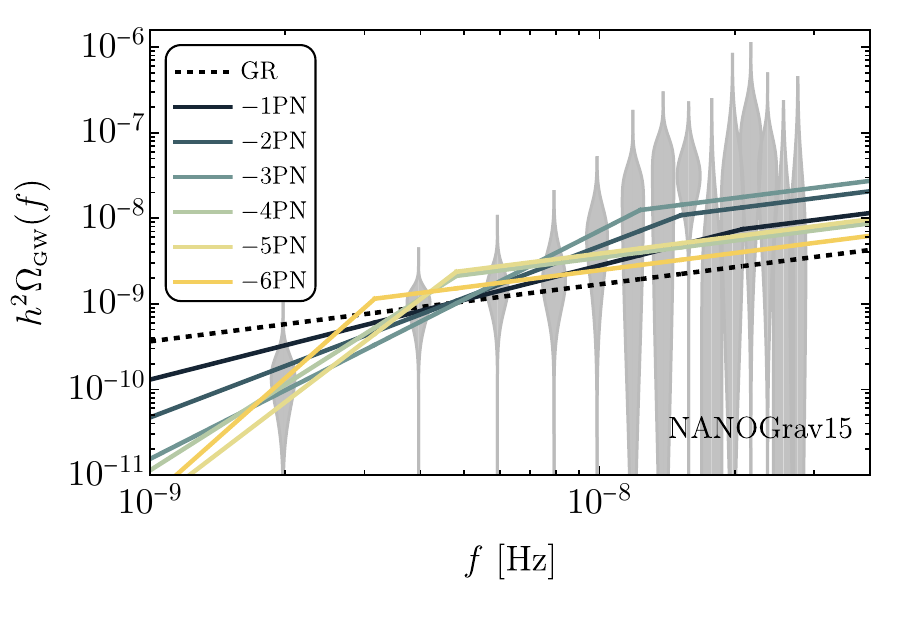}
\caption{The posterior distribution for $\Omega_\text{GW}$ in each frequency bin $f_i$ reported by NANOGrav, including HD correlations (gray violins).
We also show the best fit spectra obtained by inserting a specific PN correction in the model, i.e. assuming a BPL spectrum with $n_T = 2/3(1-q)$ at $f<f_b$  and $n_T = 2/3$ at $f>f_b$. The black dashed line shows the best fit assuming GR scaling $\Omega_{\rm GW}\propto f^{2/3}$.}
\label{fig:NG15}
\end{figure}

\begin{figure*}[t]
\centering
\includegraphics[width=0.49\textwidth]{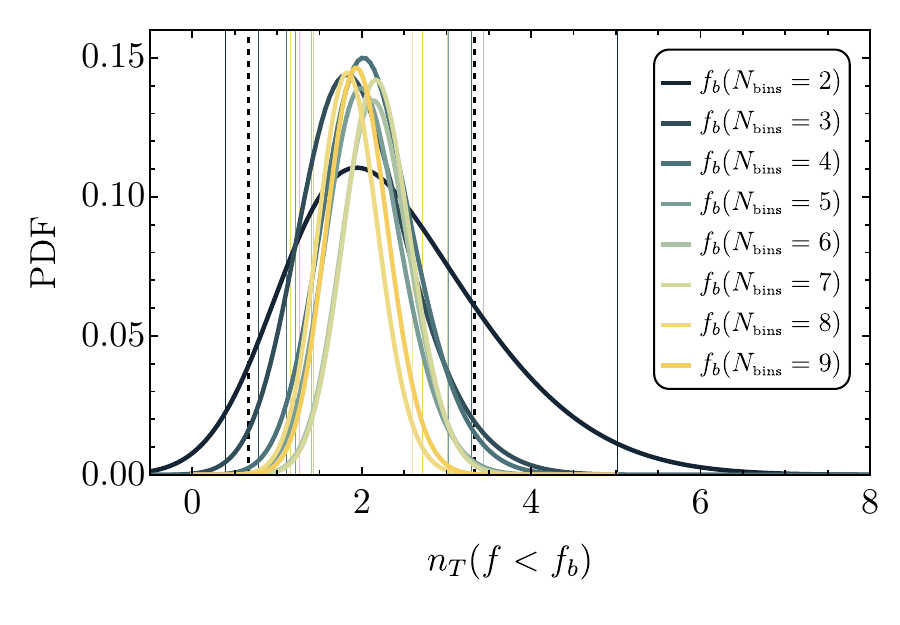}
\includegraphics[width=0.49\textwidth]{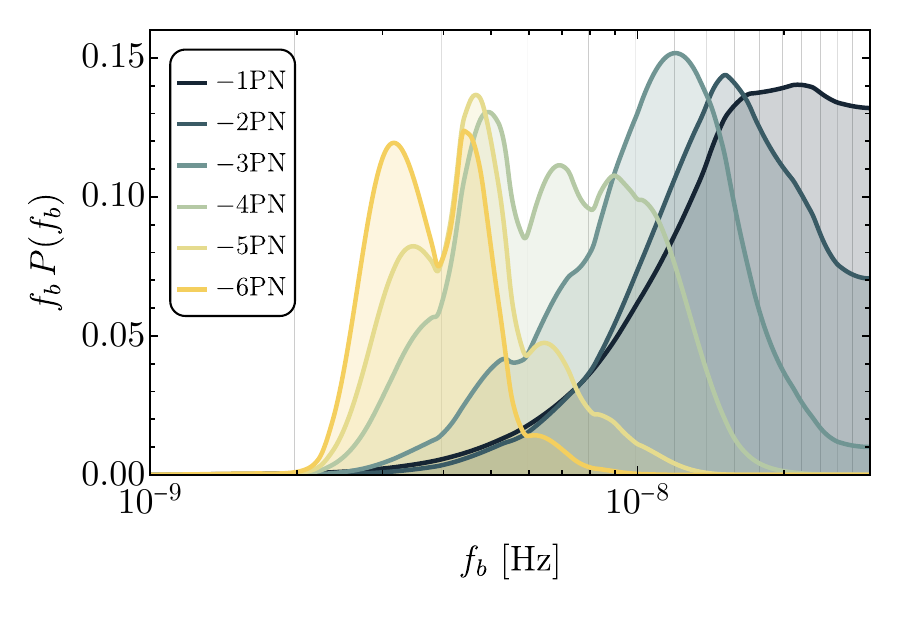}
\caption{
Fit of the NANOGrav15 dataset, assuming a BPL model with break frequency at $f_b$.
For $f>f_b$ we always assume $n_T = 2/3$ as predicted by GW-driven, circular SMBH binaries in GR. 
{\it Left:} 
The low-frequency tilt $n_T$ is left free to vary. As $f_b$ increases, the constraint becomes more stringent, and converges towards the single power-law result. The colored vertical lines indicate the 2$\sigma$ range for $n_T(f<f_b)$ while the dashed vertical lines indicate $n_T = 2/3$ and $10/3$ corresponding to GR and $-4{\rm PN}$ corrections, respectively. 
{\it Right:} 
The low frequency tilt is fixed assuming a given negative $q$PN correction, $n_T = 2/3(1-q)$ for $q = -1, \dots, -6$.
The posterior distribution for $f_b$ indicates the preference for negative PN corrections (i.e. steeper spectra than $\propto f^{2/3}$). As a consequence, we always find lower bounds on $f_b$, while for $q\lesssim -3$ upper bounds are also obtained. The light vertical lines indicate the location of the NANOGrav frequency bins.
}
\label{fig:NG15_BPLPN}
\end{figure*}

%%%%%%%%%%%%%%%%%%%%%%%%%%%%%%%%%%%%%%%%%%%
\section{Confrontation with NANOGrav15 dataset}
After the heuristic analysis above, we now focus on the most constraining PTA dataset, which is the one reported by the NANOGrav collaboration~\cite{NG15-SGWB}. 
Assuming the signal to be a power-law~(PL), this corresponds to a fraction energy density in GW today as~\cite{Allen:1997ad}
\begin{align}\label{Omega_NG_def}
    \Omega_{\rm GW}(f) = \frac{2 \pi^2 A^2 f_{\rm yr}^2}{3H_0^2}
    \left(\frac{f}{f_{\rm yr}}\right)^{n_T}  ,
\end{align}
where $H_0$ is the current Hubble rate, $f_{\rm yr}\equiv (1~{\rm yr})^{-1}\simeq 32~{\rm nHz}$, whereas $A$ and $n_T$ are the amplitude and spectral tilt, respectively. 
In Fig.~\ref{fig:NG15}, we show the posterior distribution for the SGWB abundance $\Omega_{\rm GW}(f)$ (gray violins), obtained by the NANOGrav collaboration fitting their data with the inclusion of HD correlations, in the first 14 bins.

We perform a maximum likelihood analysis to derive bounds on the tilt of the spectrum as a function of frequency, and compare this with the circular inspiral prediction $\Omega_{\rm GW}\propto f^{2/3}$.
The full log-likelihood is computed as
\begin{equation}
    \ln {\cal L} (\bm{\theta}) = \sum_k \ln p_k \left (\Omega_{\rm GW}(f_k|\bm{\theta}) \right)\,,
\end{equation}
where $\bm{\theta}$ is the parameter vector characterising the models (e.g., amplitude and tilt for the power-law model), 
while $p_k$ is the likelihood for each frequency bin $k$. 

First, we assume that the SGWB spectrum is described by a power law in {a certain number of} the lowest frequency bins,{hereafter denoted as $N_{\rm bin}$}. We show the resulting $2\sigma$ range for the spectral tilt in Table~\ref{table:ranges} (PL case).

We can also assume the SGWB spectra to be described by a broken power-law (BPL) where the low-frequency tilt ($f<f_b$) is fixed assuming a given negative PN correction, $n_T = 2/3(1-q)$, while $\Omega_{\rm GW}\propto f^{2/3}$ for $f>f_b$.
In our model~\eqref{eq:dEdfB}, this occurs naturally: for fixed $q$ and $B$, $f_b$ corresponds to saturating Eq.~\eqref{eq:inequality}, namely $f_b=B^{-\frac{3}{2q}}/(\pi G m)$.

The corresponding 2-$\sigma$ range for $n_T$, assuming different values of $f_b$ (corresponding to the location of each NANOGrav15 bin) are also shown in Table~\ref{table:ranges}.
This second version of the bounds are slightly more stringent, as additional information is brought by considering all NANOGrav15 posteriors, while assuming the high frequency range is described by the GW prediction. 
In Fig.~\ref{fig:NG15_BPLPN} (left panel), we show the corresponding posterior distribution for $n_T(f<f_b)$, assuming the BPL model.

One can also perform a different analysis and focus on a single PN correction (i.e.\ fixing $q$) and derive the best fit parameters for the location of the break frequency $f_b$. 
The posterior distribution for $f_b$ shown in the right panel of Fig.~\ref{fig:NG15_BPLPN} has always significant support in the NANOGrav band, thus indicating the preference for negative PN corrections (i.e.\ steeper spectra than $f^{2/3}$). In all cases, we find lower bounds on $f_b$, while upper bounds are also obtained when $q\lesssim -3$.

{
\renewcommand{\arraystretch}{1.4}
\setlength{\tabcolsep}{6pt}
\begin{table}[t]
\centering\begin{tabular}{c|c|cc}
\hline\hline
frequency $[{\rm nHz}]$ & $N_\text{\tiny bin}$ &  $n_T^{\text{\tiny (PL)}}$ & $n_T^{\text{\tiny (BPL)}}$   \\
\hline
% $f_2 =3.96$ & $2$ & [0.4, 5.0]  & [0.5, 5.9]  \\
% $f_3 =5.93$ & $3$ & [0.8, 3.4]  & [0.9, 4.0]  \\
% $f_4 =7.92$ & $4$ & [1.1, 3.3]  & [0.9, 3.4]  \\
% $f_5 =9.88$ & $5$ & [1.2, 3.0]  & [1.0, 3.3]  \\
% $f_6 =11.9$ & $6$ & [1.4, 3.0]  & [1.0, 3.3]  \\
% $f_7 =13.9$ & $7$ & [1.4, 3.0]  & [1.0, 3.3]  \\
% $f_8 =15.8$ & $8$ & [1.2, 2.6]  & [1.6, 3.4]  \\
% $f_9 =17.8$ & $9$ & [1.3, 2.7]  & [1.4, 3.1]  \\
$f_2 =3.96$ & $2$ & $ 3.0^{+2.9}_{-2.4} $ & $2.2^{+2.8}_{-1.8}$  \\
$f_3 =5.93$ & $3$ & $ 2.3^{+1.7}_{-1.5} $ & $1.9^{+1.5}_{-1.1}$  \\
$f_4 =7.92$ & $4$ & $ 2.0^{+1.4}_{-1.2} $ & $2.1^{+1.2}_{-1.0}$  \\
$f_5 =9.88$ & $5$ & $2.1^{+1.2}_{-1.1}$ & $2.0^{+0.1}_{-0.8}$  \\
$f_6 =11.9$ & $6$ & $2.1^{+1.2}_{-1.1}$ & $2.2^{+0.9}_{-0.8}$  \\
$f_7 =13.9$ & $7$ & $2.1^{+1.2}_{-1.1}$ & $2.2^{+0.8}_{-0.8}$  \\
$f_8 =15.8$ & $8$ & $2.4^{+1.0}_{-0.9}$ & $1.8^{+0.8}_{-0.7}$  \\
$f_9 =17.8$ & $9$ & $2.2^{+0.9}_{-0.8}$ & $2.0^{+0.8}_{-0.7}$  \\
\hline\hline
\end{tabular} %
\caption{We report the range of $n_T$ at 2$\sigma$ obtained 
fitting the first $N_\text{\tiny bin}$ frequency bins with a power-law~(PL), or with a broken power-law (BPL), where we set $n_T = 2/3$ only at $f_i > f_b$. The first frequency bin is at  $f_1 = 1.98\,{\rm nHz}$.}
\label{table:ranges}
\end{table}
}

\begin{figure}[t]
\centering
\includegraphics[width=0.9\textwidth]{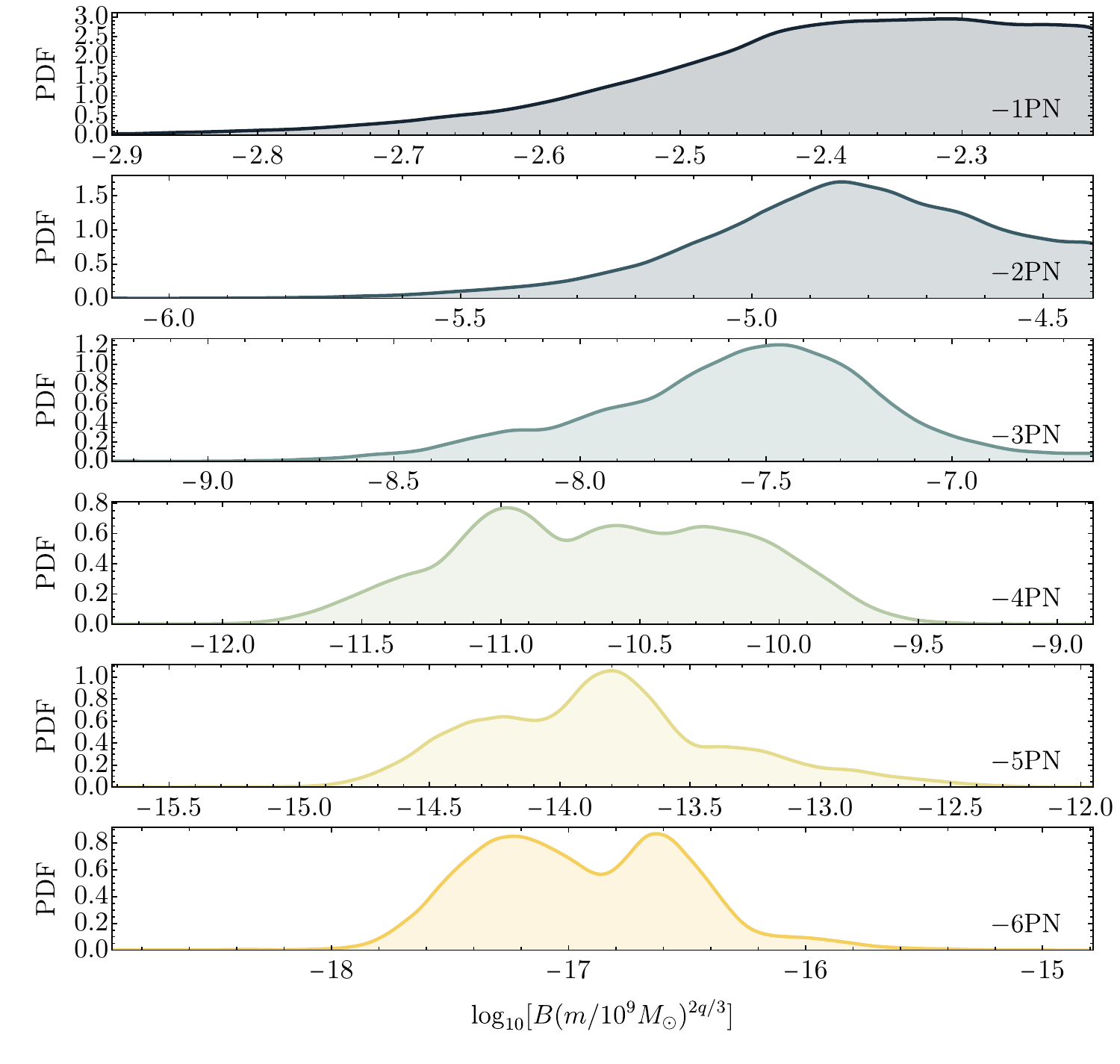}
\caption{Posterior distribution of $B$ derived assuming only one PN correction is active at the time. 
We left the explicit dependence on the total mass of the binaries, conservatively normalised to the value of $m=10^9 M_\odot$. 
These constraints are derived from the BPL posteriors on $f_b$ in Fig.~\ref{fig:NG15_BPLPN}.}
\label{fig:Bbounds}
\end{figure}

Constraints on $f_b$ can be directly translated into constraints on the parameter $B$ using Eq.~\eqref{eq:inequality}. 
We show this in Fig.~\ref{fig:Bbounds}, assuming for simplicity that all binaries have the same total mass. 
The absence of support for $B=0$ in all cases with $q<0$ is due to the existing tension in PTA data with the standard scenario $\Omega_{\rm GW}\propto f^{2/3}.$

\begin{figure}[t]
\centering
\includegraphics[width=0.9\textwidth]{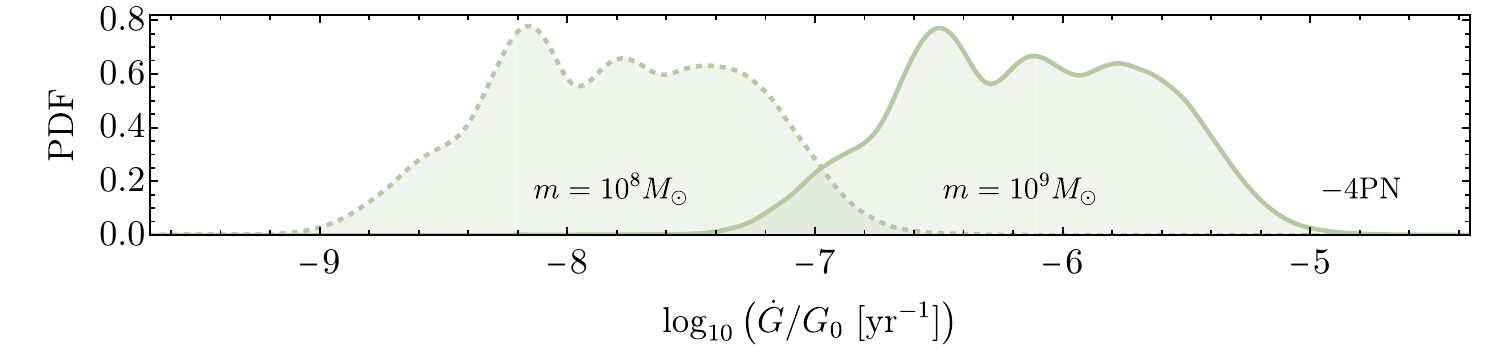}
\caption{Posterior distribution of $\dot G/G_0$ assuming two values for the binary mass, $m=10^9 M_\odot$ and $m=10^8 M_\odot $.}
\label{fig:BboundsdotG}
\end{figure}

For $q=-4$ the posterior on $B$ can be mapped into a bound on $\Dot G/G_0$, {assuming the latter to be approximately constant at the source's redshift $z\sim0.5$ dominating the signal~\cite{NG15-SMBHB}}, see Fig.~\ref{fig:BboundsdotG}.
Even in the most conservative scenario ($m=10^9 M_\odot$), the bound on $\Dot G/G_0$  is comparable to the projected bounds that LISA is expected to place using quasi-monochromatic sources~\cite{Barbieri:2022zge}.
Due to the $\Dot G\propto m^{5/3}$ dependence, in more optimistic scenarios ($m<10^9 M_\odot$) the bound is only slightly less stringent than what will be achievable in the LISA era by detecting SMBH coalescences and extreme mass-ratio inspirals~\cite{Yunes:2009bv,Chamberlain:2017fjl,Perkins:2020tra} at similar redshift ($z\approx [0.1\div1]$). Overall, the bound on $\Dot G/G_0$ is also several orders of magnitude better than the current bounds with LIGO/Virgo black-hole\footnote{Neutron-star binary GW170817 sets a bound $\Dot G/G_0\lesssim 10^{-8}\,{\rm yr}^{-1}$~\cite{Vijaykumar:2020nzc}, which anyway refers to much smaller redshift, $z\approx0.01$.} binaries~\cite{Yunes:2016jcc,LIGOScientific:2021sio}, consistently with the fact that a small-velocity nonperturbative modification in the nHz band can become perturbatively small in the relativistic regime.

%%%%%%%%%%%%%%%%%%%%%%%%%%%%%%%%%%%%%%%%%%%
%%%%%%%%%%%%%%%%%%%%%%%%%%%%%%%%%%%%%%%%%%%

\section{Conclusions and outlook} 
We argued that the recent groundbreaking PTA detection of a SGWB provides novel avenues to test gravity in the slow-motion regime.
The novel constraints derived here are already competitive with those that the future LISA mission will place using nonrelativistic binaries.
More accurate and stringent constraints should be derived by taking into account the mass distribution of SMBH binaries and its uncertainties.
Furthermore, we expect that the upcoming joint analysis involving all collaborations within the International PTA framework will soon strengthen the constraints derived in this work. Likewise, the PTA posteriors on the tilt will shrink in time as longer duration datasets will be analyzed. Following the estimates provided in Ref.~\cite{NANOGrav:2020spf}, one may expect the uncertainty on the spectral tilt to scale with the observation time as $\sigma_{n_T} \propto T_\text{obs}^{-0.9 (5-n_T)}$ ($\propto T_\text{obs}^{-0.4}$) in the intermediate (weak) signal regime, with additional improvements given by the growing number of observed pulsars. 

Depending on the specific modified gravity theory, binary pulsars~\cite{PhysRevD.101.104011,Berti:2015itd} can provide more stringent constraints on negative PN terms than PTAs. However, a direct comparison is challenging, since the coupling $B$ can be source-dependent, for example being zero for pulsars and nonzero for SMBHs, as it happens in several theories for black-hole dipolar emission~\cite{Barausse:2016eii,PhysRevD.101.104011}.
As expected, PTA constraints are more stringent for more negative PN terms. For example, even if a $-1{\rm PN}$ modification were excluded at each frequency bin, the bounds derived from Eq.~\eqref{eq:inequality} at the nHz scale for $q=-1$ would be at most the level of $B\lesssim 6\times 10^{-4}$. This is more stringent than those that can be obtained with LIGO at design sensitivity, but much less stringent than the projected bounds on $B$ with LISA sources~\cite{Barausse:2016eii,Chamberlain:2017fjl,Perkins:2020tra}.

In addition to placing competitive constraints on modified gravity, a more ambitious possibility is to invoke slow-motion, beyond-GR effects to alleviate or solve the tension in the latest PTA data if the detected SGWB is interpreted as arising from SMBH binaries. 
While we showed that this is in principle possible without violating current constraints, one should be careful with degeneracies with astrophysical effects~\cite{Sesana:2008mz,2010arXiv1002.0584K,Kelley:2016gse,Perrodin:2017bxr,Ellis:2023owy,NG15-SMBHB,NG15-NP, Bonetti:2017lnj}, including large eccentricity~\cite{Enoki:2006kj,Chen:2016zyo}, and
environmental modifications such as stellar scattering, interaction with circumbinary gas, and dynamical friction~\cite{2010arXiv1002.0584K,Sampson:2015ada,Ghoshal:2023fhh}.
In this context, since our framework can accommodate both beyond-GR and environmental effects, it can be used to disentangle these modifications (which often enter at different negative PN order) through dedicated Bayesian inferences with more constraining future datasets.
Finally, future work can be devoted to confront our general model with the hypothesis that the signal is due to, or contaminated by, sources of cosmological origin~\cite{NG15-NP}.

\acknowledgments

We are grateful to Emanuele Berti and Laura Sberna for interesting discussion.
We acknowledge the financial support provided under the European Union's H2020 ERC, Starting Grant agreement no.~DarkGRA--757480 and under the MIUR PRIN programme, and support from the Amaldi Research Center funded by the MIUR program ``Dipartimento di Eccellenza" (CUP:~B81I18001170001). This work was supported by the EU Horizon 2020 Research and Innovation Programme under the Marie Sklodowska-Curie Grant Agreement No. 101007855 and additional financial support provided by Sapienza, ``Progetti per Avvio alla Ricerca", protocol number AR2221816C515921 and AR1221816BB60BDE.

\bibliographystyle{JHEP}
\bibliography{main}

\end{document}